\documentclass[twocolumn,nofootinbib,preprintnumbers,amsmath,amssymb,superscriptaddress,prl]{revtex4}

\usepackage[pdftex]{graphicx}    
\usepackage{color}
\usepackage{slashed}
\usepackage[normalem]{ulem}
\newcommand{\ts}{{\theta}}
\newcommand{\tb}{{\overline \theta}}
\newcommand{\tu}{{\underline \theta}}
\newcommand{\cb}{{\overline c}}

\newcommand{\Tr}{\text{Tr }}
\newcommand{\UV}{\text{UV}}
\newcommand{\QCD}{\text{QCD}}
\newcommand{\Larkin}{\text{Larkin}}

\begin{document}

\title{Gribov copies, avalanches and dynamic generation of a gluon mass}

\author{Matthieu Tissier}%
\affiliation{%
Laboratoire de Physique Th\'eorique de la Mati\`ere Condens\'ee, UPMC, CNRS UMR 7600, Sorbonne
Universit\'es, 4 Place Jussieu,75252 Paris Cedex 05, France.}

\date{\today}

\begin{abstract}
  Analytic calculations in the infrared regime of nonabelian gauge
  theories are hampered by the presence of Gribov copies which results
  in some ambiguity in the gauge-fixing procedure. This problem shares
  strong similarities with the issue of finding the true ground state
  among a large number of metastable states, a typical situation in
  the field of statistical physics of disordered systems. Building on
  this analogy, we propose a new gauge-fixing procedure which, we
  argue, makes more explicit the influence of the Gribov copies. A
  1-loop calculation shows that the dynamics of these copies can lead
  to the spontaneous generation of a gauge-dependent gluon mass.

\end{abstract}

\maketitle

{\bf Introduction.} Gauge invariance is a very important concept of
modern theoretical physics which has been heavily used in the past to
build and classify new theories. Nowadays, as is well known, all
fundamental interactions (gravity, strong and electroweak
interactions) involve such symmetries. However, as soon as analytic
calculations are performed, one needs to break this invariance
explicitly, through a procedure called gauge fixing. This is in
particular true in the realm of quantum field theories, where gauge
invariance leads to zero modes that make the propagator, one of the
building blocks of field theory, ill-defined.

In its standard form, gauge fixing consists in grouping field
configurations which are related one to another through a gauge
transformation in equivalence classes, called gauge orbits. All these
field configurations bear the same physical content and one can retain
only one representant per gauge orbit. This choice is done through a
gauge condition, such as the Lorenz condition $\partial_\mu A^\mu=0$
of electromagnetism (also called Landau gauge). In quantum field
theory, gauge fixing is generally implemented by restricting the path
integral to the subset of field configurations that fulfill the gauge
condition. Once the gauge is fixed, the propagator is well defined and
one can implement the standard techniques of field theory
(perturbation theory, renormalization group, etc.) to access the
physics of the system.

However, as first understood by Gribov \cite{Gribov:1977wm} (see also
\cite{Singer:1978dk}), this procedure fails for nonabelian gauge
theories. Indeed, there typically exists several field configurations
which are equivalent up to a gauge transformation and which satisfy
the same gauge condition. These many intersections between a gauge
orbit and the gauge condition, called Gribov copies, spoil the
standard gauge-fixing procedure described above. In fact, Neuberger
\cite{Neuberger:1986xz} showed that the textbook Faddeev-Popov
gauge-fixing procedure, which overlooks the existence of Gribov copies
is actually nonperturbatively ill-defined. In principle, this result
invalidates the calculations based on the Faddeev-Popov procedure, at
least in the long-distance regime. However, very few methods are
available to overcome the Gribov ambiguity. (Even the Gribov-Zwanziger
approach, which was built to overcome this problem
\cite{Gribov:1977wm,zwanziger}, presents some residual ambiguities
\cite{vanbaal}.) This probably explains why no consensus has been
reached on the influence of the Gribov issue on actual
calculations.

There is however a consensus in the community on one unambiguous way
of treating the Gribov ambiguity in lattice simulations. This was
heavily used in the past to obtain correlation functions in the Landau
gauge, in the quenched approximation
\cite{Cucchieri:2007rg,Bogolubsky:2009dc,Dudal:2010tf,Cucchieri:2008qm}
as well as in full QCD \cite{Bowman:2005vx,Skullerud:2003qu}. We now
briefly describe this method in the context of Euclidean Yang-Mills
theory with a SU(2) group. Let us introduce the gauge field
$A_\mu= A_\mu^a \sigma^a/2$ where the $\sigma$ are the Pauli matrices
and a sum over repeated indices is implied. The gauge transformation
acts on the gauge field as
\begin{equation}
  A_\mu^U=U(A_\mu+ i{g_0}^{-1}\, \partial_\mu) U^\dagger
\end{equation}
where $U(x)$ belongs to the SU(2) group and $g_0$ is the (bare)
coupling constant of the theory.  It is easy to show that an extremum
$U^\star$ of
\begin{equation}
  \label{eq_min}
f_A[U]=  \int d^4x\ \Tr A^U\cdot A^U
\end{equation}
satisfies the Landau condition $\partial_\mu A_\mu^{U^\star}=0$. This
observation can be used to implement in a numerically efficient way
the Landau gauge on the lattice: Instead of solving the constrained
problem $\partial_\mu A_\mu^{U^\star}=0$, it proves easier to search
for a local minimum of $f_A[U]$. It is found that there typically
exist many local minima of the functional $f_A$, which is the
signature of the Gribov problem on the lattice. 

The most striking result obtained by this method concerns the
gluon-gluon correlation function which is found to saturate at low
momentum. This is surprizing because the gluons are massless in
Yang-Mills theory, which would {\em a priori} lead to a divergence of
the correlation function at low momentum. Again, no consensus has been
reached concerning the origin of this ``massive''
behavior. Calculations based on Dyson-Schwinger equations
\cite{Aguilar:2004sw,Boucaud:2006if,Aguilar:2008xm} and functional
renormalization group \cite{Fischer:2008uz} now indicate that the mass
is generated through nouperturbative effects while the refined
Gribov-Zwanziger approach relates this phenomenon to the appearence of
a condensate \cite{Dudal:2008sp}. In any case, this massive behavior
is a welcome feature because it tends to regularize the infrared
behavior of the theory. In particular, it was found that adding a mass
term to the gluon on phenomenological ground leads to infrared-safe
renormalization-group trajectories \cite{Tissier:2011ey}. This opens
the way to perturbative calculations of the low-momentum sector of QCD
correlation functions, in good agreement with those measured in
lattice
calculations~\cite{Tissier:2010ts,Tissier:2011ey,Pelaez:2013cpa}.

We can gain new intuition on the Gribov issue by considering {\em a
  priori} unrelated models of statistical physics in the presence of
quenched disordered, such as random manifolds \cite{Mezard,Bray} or
random field \cite{Imry:1975zz} problems. These models share the
property that their long-distance physics are governed by
low-temperature properties. Indeed, thermal fluctuations are found to
be negligible as compared to the sample-to-sample fluctuations of the
quenched disorder. This shows up in the renormalization-group analysis
because the fixed points that govern the critical properties are found
at vanishing running temperature. As a consequence, extracting the
critical properties of these systems requires to characterize their
ground state. This proves difficult because there exists typically a
very large number of local minima, with almost degenerate energies,
very much like the extrema of $f_A[U]$.

It is now well established that the existence of the many local
extrema is a fundamental ingredient to understand the physics of these
disordered systems. Let us briefly discuss this point in the case of
the Random-Field Ising Model (RFIM) (see \cite{Tissier:2011zz,Tissier:2011mu,Tissier:2011mv} for more details). In order to characterize the
properties of the ground state, a field-theoretical description was
proposed by Parisi and Sourlas \cite{Parisi:1979ka} that closely
resembles the Faddeev-Popov construction. It naturally leads to the
famous property of dimensional reduction \cite{dr} (that is, the
critical exponents of the RFIM in $d$ space dimensions are equal to
those of the {\em pure} Ising model in $d-2$ dimensions) to all orders
in perturbation theory, a property which is however known to be wrong
in three dimensions \cite{drbreaking}. Why does the Parisi-Sourlas
construction fail to reproduce the breaking of dimensional reduction?
Because, as the Faddeev-Popov construction, it does not properly take
into account the many metastable states of the system \cite{parisi},
and in particular the following physical effect. Take one sample of
the system, and submit it to a small external magnetic field. Since
the hamiltonian presents many local minima, the ground state is very
sensitive to the external field and there occur discontinuities in the
magnetization when the external field is changed. These collective
spin flips, called avalanches \cite{avalanches}, can involve a large
number of spins. It was shown in \cite{tarjus} that the dimensional
reduction property is violated if the {\em strength} of these
avalanches is large enough. (Here, the strength is related to the
Hausdorff dimension of spanning avalanches at criticality.) This
implies that, if the strength of avalanches is small enough, although
not justified because of the many metastable states, the
Parisi-Sourlas construction leads to the correct prediction of
dimensional reduction.

The analogy between disordered systems and the gauge-fixing procedure
relies on the fact that, in both cases, one has to face a large number
of extrema. From our knowledge on disordered system, it may or may not
be true that the Faddeev-Popov construction is a good starting point
for describing the long-distance physics of QCD. This strongly depends
on the properties of the Gribov copies, in particular on the ``size''
of the avalanches that take place when the system jumps from one Gribov
copy to another. The aim of this letter is to study this issue.

{\bf Gauge fixing.} To avoid the Gribov-Singer no-go theorem, we
follow the idea of \cite{Serreau:2012cg} of modifying the way the
gauge-fixing is implemented. Instead of trying to pick-up just one
representant per gauge orbit, we sum instead over {\em all} Gribov
copies and weight their contributions. More precisely, we fix the
gauge by looking for extrema of
\begin{equation}
  \label{eq_min_2}
  f_{A,\eta}[U]=\int \Tr (A^U A^U+\eta^ \dagger U+U^\dagger \eta)
\end{equation}
where $\eta$ is an external field, similar to the one introduced to
generate the linear gauge.  For an operator $\mathcal O[A]$, the
gauge-fixing procedure (that we note with brackets) reads:
\begin{equation}
  \label{eq_gf}
  \langle \mathcal O[A]\rangle=\frac{\sum_i s(i)\mathcal
    O[A^{U_i}]e^{-\rho_0 f_{A,\eta}[U_i]}}{\sum_i s(i) e^{-\rho_0 f_{A,\eta}[U_i]}}
\end{equation}
where the sums run over all the Gribov copies, $\rho_0$ is a gauge
parameter (of mass dimension 2), $s(i)=1$ if $f_{A,\eta}$ has an even
number of unstable directions around $U_i$, and $s(i)=-1$
otherwise. 

After fixing the gauge, we average the gluon field $A$ with the
Yang-Mills action,
\begin{equation}
  \label{eq:YM}
  S_{\text{YM}}=\int_x \frac 12 \Tr(\partial_\mu A_\nu-\partial_\mu A_\nu+g_0[A_\mu,A_\nu])^2,
\end{equation}
and the $\eta$ field with a gaussian distribution
$ \mathcal P[\eta]\propto \exp[{-{g_0^2}/{{\xi_0}}\int_x \Tr
  (\eta^\dagger\eta)}]$.  A similar procedure was used in
\cite{Serreau:2013ila}. The difference is that we now average
simultaneously over $\eta$ and $A$. In the language of statistical
mechanics, the $\eta$ fields is a quenched disorder while it was
annealed in \cite{Serreau:2013ila}.

To rewrite the corresponding action in a manageable form, a few
technical steps must be performed (see \cite{Serreau:2012cg} for more
details). First, we rewrite the numerator and denominator of
Eq.~(\ref{eq_gf}) in terms of a field theory with a matrix field $U$,
a Lagrange multiplier and a pair of ghost-antighost fields. All these
fields nicely rearrange in a SU(2) superfield
$\mathcal V(x,\theta,\bar\theta)$ where $\theta$ and $\bar \theta$ are
(anticommuting) Grassmann variables. Second, we use the replica trick
to take care of the denominator of Eq.~(\ref{eq_gf}). This amounts to
introducing $n-1$ replica of the superfield:
$\{\mathcal V_i, i\in\{2,\cdots,n\}\}$ and taking the limit $n\to 0$
at the end of the calculation. Third, we factorize the gauge group by
integrating over one (out of $n$) matrix fields $U$. We end up with a
local action that can be decomposed according to the number of replica
sums, $S=S_{\text{YM}}+S_0+S_1+S_2$, with
\begin{align}
  \label{eq:S0}
  &S_0=\int_x \frac 12(D_\mu\cb^a\partial_\mu
  c^a+\partial_\mu\cb^a D_\mu c^a)
+h^a\partial_\mu
  A_\mu^a\\
\nonumber&-\frac {\xi_0}2 (h^a)^2-\frac {g_0^2\xi_0}4(\epsilon^{abc}\cb^b c^c)^2   +\frac{{\rho_0}}2(A_\mu^a)^2 +{\rho_0\xi_0} \cb^a c^a\\
  \nonumber
    &S_1=\sum_{i=2}^n\int_{x\tu_i}\Tr (A_\mu-\frac i {g_0}\mathcal
      V_i^\dagger\partial_\mu\mathcal V_i)^2\\&\qquad+{{\xi_0}}(n_i^0\cb^ac^a-\frac 4
  {g_0^2}n_i^0\rho_0+\frac{{2}}{g_0}n_i^ah^a)\nonumber\\
  \label{eq:S2}
  &S_2=-\frac{{2\xi_0}}{g_0^2}\sum_{i,j=2}^n\int_{x\tu_i\tu_j} (n_i^0n_j^0+n_i^an_j^a)
\end{align}
where we have introduced the representation of SU(2) matrices in terms
of 4-components unit-vectors
$\mathcal V_i=n_i^0 \openone+i \sigma^a n_i^a$.  $S_0$ is the
Curci-Ferrari action \cite{Curci76}. Each replicated field
$\mathcal V_i$ (or $n_i$) in $S_1$ and $S_2$ comes with its own set of
Grassmann variables which are integrated over with a nontrivial
measure: $\int_{\tu}=\int d\ts d\tb(\rho_0 \tb \ts-1)$
\cite{Tissier:2008nw}. The two-replica action Eq.~(\ref{eq:S2}) is of
particular importance in what follows because this is where the
dynamics of the aforementioned avalanches is incoded. This term
originates from the quenched average over the $\eta$ field and was not
present in the similar implementation of \cite{Serreau:2013ila}.

{\bf 1-loop calculation.} We now want to perform a perturbative
analysis of this action. However, it proves important to first
determine the general structure of the divergent part of the average
action $\Gamma$. This can be achieved by using the Ward (or
Slavnov-Taylor) identities associated with the symmetries of the bare
action. Apart from the obvious invariance under translation, rotations
and rotations in color space, the action is invariant under a BRST
transformation:
\begin{equation}
  \begin{split}
  \label{eq:BRS}
  &sA_\mu^a=\partial_\mu c^a+g_0\epsilon^{abc} A_\mu^b c^c\qquad sc^a=-\frac {g_0}2 \epsilon^{abc}c^b c^c\\&
            s\cb^a=h^a -\frac{{g_0}}{2} \epsilon^{abc}\bar c^b c^c
            \qquad s\mathcal V_i=-i\frac{g_0}2\mathcal V_i c^a\sigma^a\\
&s h^a={\rho_0} c^a+\frac{{g_0}}{2} \epsilon^{abc} \Big(h^b 
  c^c+\frac{{g_0}}{4} \epsilon^{cde} \cb^b  c^d c^e\Big)
\end{split}
\end{equation}
and a similar anti-BRST transformation. Other constraints can be
obtained by considering a transformation acting simultaneously on all
superfields:
\begin{equation}
  \delta^a \mathcal V_i=i\sigma^a \mathcal V_i
\end{equation}
This transformation is not a symmetry, but the variation of the action
can be expressed in terms of the fields and their BRST variations. We
can thus deduce Slavnov-Taylor identities. Finally, we can find strong
constraints by performing a shift of the $h$ field in the partition
function, which implies a local Ward identity.

By using all these constraints, a straightforward but lengthy
calculation shows that the microscopic action is not
renormalizable. It can however be made renormalizable (up to some
reparametrization of the field $\mathcal V$) by considering more
general forms for the 1 and 2 replica parts:
\begin{align}
  \label{eq:S1t}
    \tilde S_1=&v_0\sum_{i=2}^n\int_{x\tu_i}\Tr (A_\mu-\frac i {g_0}\mathcal
      V_i^\dagger\partial_\mu\mathcal V_i)^2\\&
    -\frac{4\xi_0\rho_0}{g_0^2} W_0(n^0_i)- {\xi_0}W_0''(n^0_i)(\cb^a n_i^a c^b n_i^b)\nonumber\\
    &+\xi_0 W_0'(n^0_i)\left (n_i^0\cb^a c^a+\frac 2{g_0} n_i^ah^a\right)\nonumber\\
  \label{eq:S2t}
  \tilde S_2&=-\frac{{2\xi_0 v^2}}{g_0^2}\sum_{i,j=2}^n\int_{x\tu_i\tu_j}R_0(n_i^0n_j^0+n_i^an_j^a)
\end{align}
where $W_0$ and $R_0$ are arbitrary functions. We recover the
original action for
\begin{equation}
\label{eq_init_cond}
  v_0=1,\qquad R_0(x)=W_0(x)=x.
\end{equation}
We have checked that the action obtained by choosing $v_0=1$ and
$R_0=W_0$ a generic function can be interpreted as a gauge-fixed
action for a constraint more general than that of (\ref{eq_min_2}).

The one-loop calculation of the divergent part of the effective action
is now rather standard except maybe for the fact that we need to
renormalize whole functions ($R$ and $W$). This is a usual feature of
disordered systems and the most adapted method consists in performing
the calculations in a nontrivial background for the fields $n_i$. This
is clearly efficient because we can extract the renormalization of the
whole $R$ and $W$ functions in a single calculation. It also gives
nontrivial tests of the loop calculations because a same
renormalization factor can be extracted by different means. We have
checked that all these definitions lead to the same renormalization
factors in our calculation.

The one-loop calculation leads, in the Minimal Scheme, to the following
$\beta$ functions: 
\begin{align}
\label{eq_beta_u}
  &\beta_u=-\frac{22}3 u^2\qquad\qquad\qquad\quad\beta_\xi= \xi \frac u 6(26-3\xi)\\
\label{eq_beta_rho}
&\beta_\rho= \rho \frac u 6(-35+3\xi)\qquad \qquad\beta_v=v \frac{u\xi} 2[R'(1)-1]\\
\label{eq_beta_R}
&\beta_R(x)=\nonumber u \frac \xi8\Big\{4R(x)-8R(x)R'(1)\\&\qquad+(1+2x^2)R'(1)^2-2[R'(x)-xR'(1)]^2\\&\nonumber\qquad-[R'(1)-xR'(x)+(1-x^2)R''(x)]^2\Big\}
\end{align}
where the interaction is expressed in terms of
$u=g^2/(8\pi^2)=\alpha/(2\pi)$. We retrieve the Curci-Ferrari $\beta$
functions \cite{Curci76} for the coupling constant $u$, gauge
parameter $\xi$ and gluon mass $\rho$. Finally, the flow of $R$
closely resembles that of the two-replica function $R$ in the random
field O(4) model \cite{fisher}, except for the multiplicative factor
and the first term in braces. This is not a surprize because, as
already mentioned, the SU(2) matrices can be parametrized in terms of
a 4-component unit vector, which yields a nonlinear representation of
the O(4) group. We observe that the function $W$ does not appear in
the flow equations for the other parameters and we therefore omit it
in the remaining of the Letter.

{\bf Integration of the RG flow} Let us now study the
renormalization-group flow and its consequences for the theory under
study. We initialize the flow at some ultraviolet scale $\Lambda_\UV$
which would corespond to the inverse lattice spacing of a lattice
simulation with some initial value of the coupling constant $g_\UV$,
gauge parameter $\xi_\UV$ and weight $\rho_\UV$. The flows of $u$,
$\xi$ and $\rho$ decouple from the rest and we can integrate
analytically these expressions. As usual, $u$ presents a Landau pole
at a scale $\Lambda_\QCD=\Lambda_{\UV}\, e^{-3/(22 u_\UV)}$.

Next, we consider the flow of $v$. It involves $R'(1)$ whose flow can be
easily deduced from Eq.~(\ref{eq_beta_R}):
\begin{equation}
  \label{eq_betaRp1}
  \beta_{R'(1)}=({u\xi}/8)R'(1)[1-R'(1)]
\end{equation}
(Note that the flow of $R'(1)$ does not involve higher derivatives of
$R$!)  We would naively deduce that the initial value $R'(1)=v=1$ [see
Eq.~(\ref{eq_init_cond})] is a fixed point. However, we should keep in
mind that we need to renormalize the whole function $R$. In fact,
experience from disordered systems shows that $R''(1)$ is of
interest \cite{fisher,2loops}. Using again Eq.~(\ref{eq_beta_R}) [and $R'(1)=1$] we find:
\begin{equation}
  \label{eq_betaRs1}
  \beta_{R''(1)}=-(u{\xi}/4)[11R''(1)^2+4R''(1)+1]
\end{equation}
Note that $\beta_R''(1) <0$ for all values of $R''(1)$ and that it
increases quadratically for large $R''(1)$. Consequently, the flow of
$R''(1)$ may eventually end up in a singularity at some finite length
scale, called Larkin scale in the context of disordered systems
\cite{Larkin}. We find that if
$26/3 >\xi_\UV >\xi_c=26/3(1-e^{[2 \arctan
  ({2}/{\sqrt{7}})-\pi]/\sqrt{7} })\simeq 4.35$, the Larkin scale is
larger than the Landau pole scale. In this regime, we find that
\begin{equation}
  \Lambda_\Larkin=\Lambda_\QCD \exp\left[{\frac3{22u_\UV}\left(1-\frac {\xi_c}{\xi_\UV}\right)^{22/13}}\right].
\end{equation}

\begin{figure}[t]
  \centering
  \includegraphics[width=.9\linewidth]{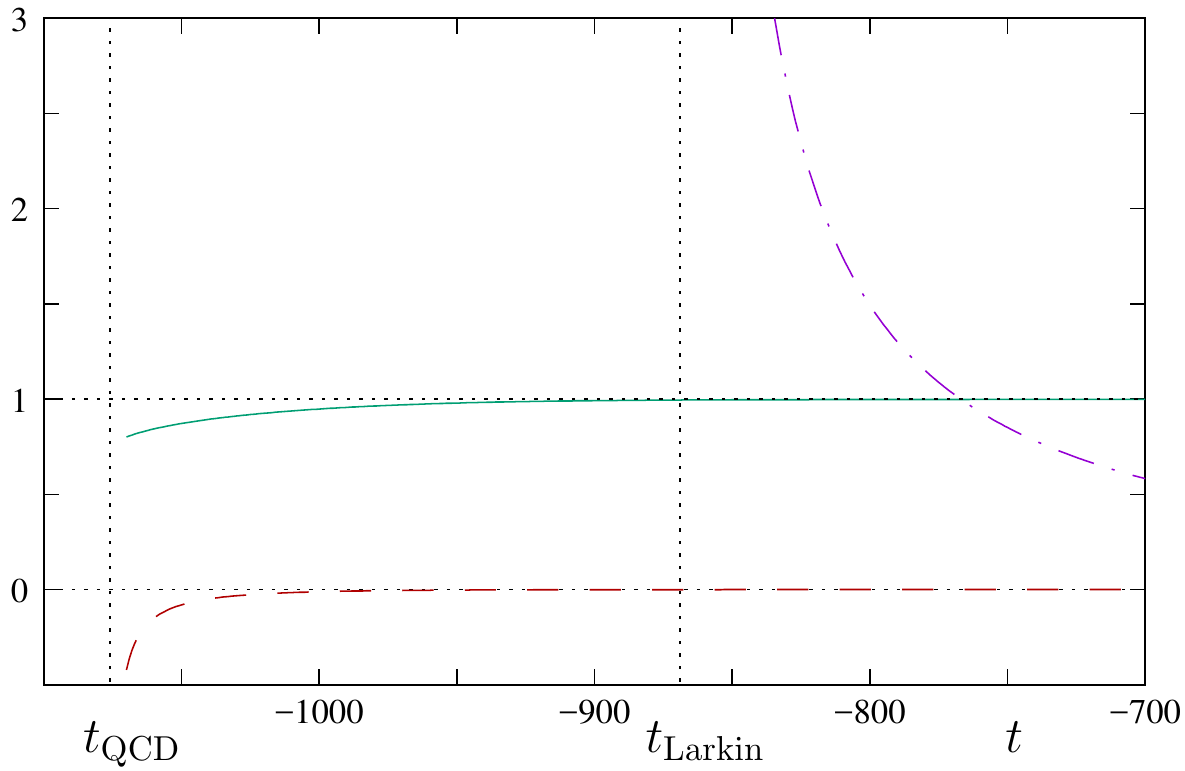}
  \caption{ Numerical integration of the flow equations with initial
    condition $g=0.1$, $\xi=7$, $\rho=0.5$ and $R'(x)=1$. $R''(1)$
    (dash-dotted line) diverges at $t=t_\Larkin$. For $t<t_\Larkin$,
    $R'(1)$ (full curve) departs from 1 and the gluon square mass
    (dashed curved, in arbitrary units) departs from zero.}
  \label{fig_Rpxt}
\end{figure}

What have we gained? It seems, at this point that the perturbative
solution disappears either at the Landau pole (for $\xi_\UV<\xi_c$) or
at an even more ultraviolet RG scale. However, the appearance of a
singularity at the Larkin scale is not as disastrous as the Landau
pole. Indeed, we can treat Eq.~(\ref{eq_beta_R}) as a partial
differential equation and go through the Larkin scale.  A careful
analysis shows that, as in the random field case \cite{2loops}, beyond
this scale, $R'(x)$ behaves as $\text{Cte}-a\sqrt {1-x}$. This spoils
the Taylor expansion which led to (\ref{eq_betaRp1}, \ref{eq_betaRs1})
and the function $R'(1)$ starts departing from 1. As a consequence,
$v$ itself starts to flow away from 1. We illustrate these behaviors
in Fig.~\ref{fig_Rpxt}.

We are now in position to discuss the generation of the gluon mass
which, looking at the renormalizable action, recieves contributions
from $S_0$ and $\tilde S_1$. In total, the ``gluon square mass'' reads
$\rho+(n-1)\rho v\to\rho(1-v)$ in the limit $n\to 0$. This implies
that, as long as the coefficient $v$ is strictly equal to 1 ({\em
  i.e.} in the range $\Lambda_\Larkin<\mu<\Lambda_\UV$), the gluon
mass vanishes, while it departs from zero at lower momenta. Since
$v>1$ in this regime, the gluon square mass is negative, which implies
that the gluon field acquires a nonzer expectation
value. 

{\bf Conclusion.} We have proposed a gauge-fixing procedure where the
Gribov ambiguity is under control. It makes possible to study the
influence of avalanches which correspond to collective changes when
the system jumps from one Gribov copy to another. A 1-loop
renormalization-group study shows that these avalanches can lead to a
singularity in the 2-replica potential, which, in turn generates a
gauge-dependent gluon mass. This scenario only work for large enough
gauge parameter $\xi$, which exclude the widely studied case of the
Landau gauge. Further investigations are needed to study whether this
gluon mass is sufficient to avoid the Landau pole, as was already
shown in the Curci-Ferrari model
\cite{Tissier:2010ts,Tissier:2011ey}. Another important study consists
in understanding the consequences of a nonvanishing expectation value
for the gauge field.

{\bf Acknowledgment} I thank U. Reinosa, J. Serreau, G. Tarjus,
N. Wschebor for useful discussions and acknowledge support of the PICS
``IRQCD''.

\end{document}